# Non-parametric Estimation approach in statistical investigation of nuclear spectra

M. A. Jafarizadeh[a,b][1], N. Fouladi[c][2], H. Sabri[c], B. Rashidian Maleki[c]

[a]Department of Theoretical Physics and Astrophysics, University of Tabriz, Tabriz 51664, Iran.

[b]Research Institute for Fundamental Sciences, Tabriz 51664, Iran.

[c]Department of Nuclear Physics, University of Tabriz, Tabriz 51664, Iran.

[*] E-mail:jafarizadeh@tabrizu.ac.ir
[2] E-mail:fouladi@tabrizu.ac.ir

## Abstract

In this paper, Kernel Density Estimation (KDE) as a non-parametric estimation method is used to investigate statistical properties of nuclear spectra. The deviation to regular or chaotic dynamics, is exhibited by closer distances to Poisson or Wigner limits respectively which evaluated by Kullback-Leibler Divergence (KLD) measure. Spectral statistics of different sequences prepared by nuclei corresponds to three dynamical symmetry limits of Interaction Boson Model(IBM), oblate and prolate nuclei and also the pairing effect on nuclear level statistics are analyzed (with pure experimental data). KD-based estimated density function, confirm previous predictions with minimum uncertainty (evaluated with Integrate Absolute Error (IAE)) in compare to Maximum Likelihood (ML)-based method. Also, the increasing of regularity degrees of spectra due to pairing effect is reveal.

## Introduction

The investigation of non-linear systems with chaotic dynamics has been considered as one of interesting topics in past decades [1-2]. Random matrix Theory (RMT) has been regarded as the most used tool in the investigation of non-fixed properties of very excited nuclei [1-5]. Different statistics such as Nearest Neighbor Spacing Distribution (NNSD) [1-12], Dyson-Mehta $\Delta_3$ statistics [2-5] and etc [3] have been applied in statistical investigations while the most common used one is NNSD or P(s). There are generally speaking, two methods are usual to deal with NNSDs, i) the histogram which is constructed by level spacing distribution (after unfolding procedure) was compared with Poisson and Wigner curves which better correspondence with one of them explore regular or chaotic dynamics respectively[6-7]. This method can't exhibit spectral statistics correspond to intermediate situation of these two limits while the interpolation between limits is very usual for different nuclear systems. ii) A least square fit of histogram with well-known distributions as Brody and etc [8-10] have been done where the estimated value for every distribution's parameter exhibit statistical properties of studied system. The great uncertainty of estimated values and also inappropriate results in some sequences with small size of data has been considered as problems of this approach. Some suggestions which based on Maximum Likelihood Estimation (MLE) [11] and Bayesian Estimation Method (BEM)[12] (other usual methods of parametric estimation methods) have been made despite their success in order to increase the accuracy of estimated values, the complicated calculations (especially BEM) make it hard to get appropriate results in all sequences. Also as other parametric estimation methods, MLE and BEM assume a particular form of the density function where one had to estimate the parameter(s) of distribution and this decrease the accuracy of estimated values.

On the other hand, histogram is one of non-parametric estimation methods which the present use of it, is subjective but some disadvantages as the numbers of bins [13-22], the position and also the bin size [14-16] affect its results. Also the lack of convergence to the right density function (if the data set is small [14]), smoothness estimators, discontinuous and etc [15-16], have been interpreted as problems which have notable effects on the precision of estimation in this approach. Different suggestions in order to overcome these problems have been occurred while the best one is Average Shifted Histogram (ASH) [18] [the average of several histograms with equal bin width but different bin location]. On the other hand, ASH can thought as a simple kernel density estimator to providing a convenient summary of a univariate set of data [18].

Kernel Density Estimation (KDE) technique prepares a non-parametric way for estimating the probability density function [17-24] independent of special distributions. Unlike histograms, even with a small number of samples, KDE leads to a smooth, continuous and differentiable density estimate. KDE does not assume any specific underlying distribution and, theoretically, the estimate can converge to any density shape with enough samples. Also, Very fast learning (simply store instances), No prior assumptions about form of model, fit any distribution with enough data, growing the complexity of estimators with amount of data and also continuous density estimator for soft kernels can be regarded as Kernel-based advantages in non-parametric estimations.

To investigate the spectral statistics of different nuclear systems independent of restrictions due to parametric estimation methods (LSF, MLE and BEM [13]) and improve the disadvantages of the other non-parametric estimation methods (Histogram and K-Nearest Neighbor estimation [14-18]) and also increase the accuracy of estimation procedure, the KDE method is used. Also, to exhibit statistical dynamics of systems, Kulback-Leibller Divergence (KLD) measure [23-24] is evaluated while the closer distance to Poisson or Wigner limits explore regular or chaotic dynamics of different sequences respectively.

To exhibit the advantages of the non-parametric estimation methods in compare to MLE method (while yields accuracies very close to CRLB in compare to other parametric estimation methods) [10-11], spectral statistics of sequences prepared by nuclei correspond to three dynamical symmetry limits in the Interaction Boson Model (IBM) [25-37] and also nuclei correspond to oblate and prolate classifications in the Bohr-Mottelson Geometric Collective Model (BMM) [33-38] frameworks (with $2^+, 4^+$ levels of pure experimental data [39-41]) is investigated. The KD-based density function and also the KLD measures confirm previous predictions with more accuracy in compare to ML-based estimated distributions(Brody and Abul-Magd) [uncertainties evaluated by Integrated Absolute Error (IAE) method which is the common technique to contrast their finite sample efficiency with the Gaussian kernels].

Also in order to investigate statistical properties due to pairing effects on nuclear spectra [42-63], we have prepared different sequences of even ($2^+, 4^+$ levels) and odd ( $\frac{1^+}{2}, \frac{3^+}{2}, \frac{5^+}{2}$ levels) mass nuclei in different mass region. Also the effect of pair types (proton-proton or neutron-neutron pairs) is analyzed while the increasing of regularity degrees of spectra in even-mass nuclei is confirmed. Furthermore, the competition between pairing and Coulomb forces yield more regular dynamics for systems with neutron-neutron pairs in compare to proton-proton pairs.

This paper is organized as follows: section 2 dealt with reviewing a statistical approach and Kernel Density Estimation (KDE) method and details about KLD and ISE, section 3 briefly summarizes the theoretical aspects of pairing effect on nuclear structure and finally, section 4 contains the numerical results obtained by applying the KDE non-parametric method to different sequences. Section 5 is devoted to comparison of KDE method with other parametric and non-parametric estimation methods, based on results given in section 4.

## 2. Statistical analysis

Statistical features of nuclear spectra have been studied with different statistics as Nearest Neighbor Spacing Distribution (NNSD) [1-12], Dyson-Mehta $\Delta_3(L)$ [4], and linear coefficients between adjacent spacing [3] which based on predictions of Random Matrix Theory (RMT). The most commonly used is NNSD or $P(s)$ functions while two main methods are usual in dealing with them, the comparison of the histogram (non-parametric estimation method) of every sequence with Poisson and Wigner curves. The better correspondence, explore regular or chaotic dynamics respectively. In addition to restriction occur by histograms (as have mentioned in introduction), this technique can't exhibit spectral statistics interpolate between these two limits.

The second method is based on parametric estimation methods included LSF, MLE and BEM [1-12] while estimates the parameter of different distribution functions as Brody [8] and etc [9-10]. The restrictions due to the nature of every distribution (the majority of them are one-parametric distributions where the resultant function even with ML estimated value is far from the exact distribution of sequence) and also complicated evaluation to estimate statistical properties (especially in BEM and MLE which estimate with more accuracy in compare to LSF) can be considered as the problems of this approach.

Different attempts have been considered to improve the disadvantages of Histogram (which based on non-parametric estimation methods [14-24]) have been considered where the best one is Average Shifted Histogram (ASH). In the simplest version of ASH, Scott [17] have been proposed to average of several histograms have same bin size but different bin location (this approach improve the dependence to the size and center of bin). On the other hand, as the number of histograms in ASH become infinity, the ASH approximate as a kernel estimate of probability density function. Kernel Density Estimation (KDE) method [19-24] regard as an alternative to the histogram, which employs kernel to smooth samples. This will prepare a smooth probability density function, which will in general more accurately reflect the underlying variable and remove the estimation dependencies of bin starting point.

- **Kernel Density Estimation (KDE)**

In statistical application, Kernel Density Estimation (KLD) is regarded as a non-parametric technique for estimating the probability density function for sequence prepared by random variables [14-24]. KDE is a fundamental data smoothing processes which inferences about the population are made based on a finite data sample. In some fields such as signal processing and econometrics it is also known as the Parzen–Rosenblatt window method [19]. An overall description about this method has presented in [14-22], but we would review important and essential concepts which are going to be used in our investigation.

The simplest form of non-parametric D.E. is the familiar histogram. The methods of choosing histogram width and also the smoothing parameter of kernel density estimators by using of data are introduced in [19]. These methods are based on estimators of risk functions corresponding to Mean Integrated Squared Error (MISE) and Kullback-Leibler information measure and other distance criteria. Assume $X_1, \ldots, X_n$ are independent, identically distributed, real valued random variables with probability density $f$. We would consider estimators $\hat{f}$ of $f$. We propose I($=I_k$) as partition of real line into disjoint intervals. If $h_k$ indicates the length of $I_k$ and $N_k = \#\{i: X_i \in I_k, 1 \leq i \leq n\}$ represents the number of observations in $I_k$, and also $X = (X_1, \ldots, X_n)$ and [19]

$$\hat{f}(x) = \hat{f}_I(x, X) = \frac{N_k}{nh_k} \qquad , \quad x \in I_k \qquad (2.1)$$

Therefore, $\hat{f} = \hat{f}_I$ is the histogram corresponding to $I$ and $h_k$. It is conventional to suppose same length for all intervals $I_k$.

The kernel estimator $\hat{f} = \hat{f}_a$ will be defined as ($a = (K, h)$) [19-22]

$$\hat{f}(x) = \hat{f}_a(x, X) = \frac{1}{nh} \sum_{i=1}^{n} K\left(\frac{x - X_i}{h}\right) , \qquad (2.2)$$

Where $K$ is a kernel, a non-negative real function which integrates to one ($\int K(x)dx = 1$). The main point in the application of this technique will be in choosing the width of bin (the smoothing parameter) and also the kernel density function by use of data. Similar to histogram case, these methods are based on estimators of risk functions corresponding to Hellinger distance (HD) or integrated squared error (ISE) which defined as follows ( For an estimator $\hat{f}$ of $g$ ) [19-22];

$$\int \left|\hat{f}(x)^{\frac{1}{2}} - g(x)^{\frac{1}{2}}\right|^2 dx \qquad Hellinger\ distance \qquad (2.3a)$$

$$\int |\hat{f}(x) - g(x)|^2 dx \qquad integrated\ square\ error \qquad (2.3b)$$

While the common measure in evaluating the efficiency of a density estimator which is based on Integrated Squared Error (ISE) [19]. We have examined different kernel functions and also different bin sizes to adopt the best density estimator and smoothing parameter which minimize ISE while Gaussian kernel [19-22]

$$K\left(\frac{x - X_i}{h}\right) = \frac{1}{\sqrt{2\pi}} e^{-\frac{\left(\frac{x-X_i}{h}\right)^2}{2}} , \qquad (2.4)$$

Yields the best efficiency (minimum ISE) with definite bin sizes (we will represent the value of bin size in the following for every sequences). To investigate chaotic or regular dynamics of nuclear spectra with KD-based estimated density function (for different sequences), we would evaluate the distances of $\hat{f}(x)$ related to Gaussian Orthogonal Ensemble (GOE) [1-12]

$$P(s) = \frac{1}{2}\pi s e^{-\frac{\pi s^2}{4}} , \qquad (2.5)$$

Where have been used in describing the chaotic properties of spectra or Poisson limit which investigate regular limit of spectra as [1-12]

$$P(s) = e^{-s} , \qquad (2.6)$$

With Kullback-Leibler Divergence (KLD) measure which is defined as[23-24]

$$D_{KL}(P\|Q) = \int P(x) \log \frac{P(x)}{Q(x)} dx \qquad (2.7)$$

KLD measure is a non-symmetric measure to exhibit the average of the logarithmic difference between the probability distributions $P(x)$ and $Q(x)$. If $D_{KL}(P\|Q) \to 0$, a closer correspondence would be appeared between two probability distribution functions, therefore, a closer distances to Poisson or GOE limits, explore regular or chaotic dynamics of sequences respectively.

## 3. Pairing Hamiltonian

In nuclear physics, pairing and quadrupole-quadrupole interactions can be regarded as the most important interactions [42-55]. This concept was proposed by Racah as a seniority scheme in atomic physics [63]. Pairing is regarded as a simple and most regular part of nuclear interaction [54-62]. In the low-lying part of nuclear spectra, it yields a pair condensate that influences strongly on all nuclear properties [42-63]. On the other hand, According to the standard BCS description borrowed from the macroscopic theory of superconductivity, the excitation of the system breaks pairs, removing them from the interaction domain and blocking the scattering phase space for remaining pairs. Then, at some excitation energy ~ or temperature a sharp second-order phase transition occurs to a normal-heated Fermi liquid where the pairing effects are usually neglected [63]. The thermodynamically properties [44-50], entropy [51-55] and etc have been studied by different authors. The statistical properties of energy levels can be used for investigating the pairing effect on nuclear structures, too. The previous descriptions of pairing [58-59] usually employs it as a one part of nuclear Hamiltonian and with variation of pairing effect, statistical properties have been analyzed in some special nuclei [51-60].

As it was mentioned in introduction, our investigations have carried out with pure experimental data but we review the role of pairing Hamiltonian briefly in order to clarify theoretical aspects of pairing (complete descriptions are presented in [42-55]).To study obvious pairing effect in Hamiltonian, some authors use the following Hamiltonian[59],

$$H = H_{real} + H_{pair}(G) \quad , \tag{3.1}$$

Where $H_{real}$ is a realistic Hamiltonian and [59]

$$H_{pair} = -G \sum_{j,j'} \sum_{m,m'>0} (-1)^{j+m}(-1)^{j'+m'} a^\dagger_{jm} a^\dagger_{j,-m} a_{j',-m'} a_{j',m'} \quad , \tag{3.2}$$

To investigate the pairing effect in nuclear spectra, they have carried their calculation with different values of *G*. Their results obviously display important effect of pairing in low-lying region of energy spectra, on the other hand, when they have assumed the exotic states in their calculation, broken pairs and phase transitions caused to neglecting the pairing effects [42-59].On the other hand, pairing force in addition to quadrupole interaction has been regarded as the main reason in constructing bosons in the interaction Boson Model (IBM) framework.. The IBM [25-30] is expressed in terms of a U(6) Lie algebra spanned by the bilinear combinations of five pairs of $L = 2$ (*d*-boson operators) and one pair of $L = 0$ (*s*-boson operators). The most general form of IBM Hamiltonian can be written in terms of s- and d-boson operators as[25-28]:

$$H = E_0 + c_0 \hat{n}_d + c_2 Q^\chi . Q^\chi + c_1 L^2 \quad , \tag{3.3}$$

$\hat{n}_d = d^\dagger . \tilde{d}$ represents the number operator of d-bosons, L is angular momentum and $Q^\chi$ is quadrupole operator defined as [25-28]:

$$Q^\chi = (d^\dagger \times \tilde{s} + s^\dagger \times \tilde{d})^2 + \chi(d^\dagger \times \tilde{d})^2, \tag{3.4}$$

with $\mathcal{X}$ as control parameter. The IBM has three dynamic symmetry limits corresponds to the following algebra chains

$$U(6) \supset \begin{Bmatrix} U(5) \supset O(5) \\ SU(3) \\ O(6) \supset O(5) \end{Bmatrix} \supset O(3) \qquad \begin{matrix} I \\ II \\ III \end{matrix} , \qquad (3.5)$$

Chain (I) describes vibrational nuclei or U(5) limit which is yield with $c_2 = 0$, chain (II) is occurred for $c_0 = 0$ & $\mathcal{X} = -\frac{\sqrt{7}}{2}$ to describe rotational nuclei or SU(3) limit and chain(III) is arises with $\mathcal{X} = 0$ & $c_0 =$ to describe γ-unstable nuclei or O(6) limit while displayed in Casten triangle [31] as Figure 1. The spectral statistics of three dynamical symmetry limits of this IBM have been considered with different methods based on parametric estimation (LSF, BEM and MLE) methods [10-11,29-30,33].

In the following, we will investigate the statistical properties of different sequences with KLD-based method to compare the accuracy of estimated density function with parametric-based estimated distributions. The distances between estimated function and exact distribution of every sequence (or the uncertainty of estimation procedure) have been obtained by Integrated Absolute Error (IAE) method which defined as[18-19]

$$\int |\hat{f}(x) - g(x)| \, dx \qquad (3.3)$$

Where $\hat{f}(x)$ is the estimated function with KD-based method or well-known distributions (we have considered Brody one as the most-common used distribution) where their parameter estimated by MLE method (as have presented in [11], ML method reduces the uncertainty of estimated values and yield estimator's variance very close to Cramer-Rao Lower Bound(CRLB) in compare to LSF and BEM methods). Also $g(x)$ represents the ratio of the number of $s_i$(level spacing in sequence while introduced in the following) $m$, to the total number of level spacing $N$

$$g(x) = \frac{m}{N} \qquad (3.4)$$

In general, the integrals (Eq.(2.7) and (3.3)) can be solved by Monte Carlo and Simpson method while we have used Simpson rule in our calculations.

## 4. Numerical results

### 4.1. Comparison the accuracy of non-parametric estimation versus parametric estimation

In order to compare the accuracy of KD-based estimated function with parametric estimation methods (ML estimated values for Brody and Abul-Magd distributions), we have prepared five sequences by nuclei corresponds with three dynamical symmetry limits in IBM and also nuclei correspond to oblate and prolate classification of BMM by pure experimental data similar to [10-11,33] (we didn't repeat this comparison with LSF or BEM estimation methods while as persisted in [11], MLE method prepares the more precision in compare to other parametric estimation methods). In order to unfold our spectrum, we had to use some levels with same symmetry [3].This requirement is equivalent with the use of levels with same total quantum number (J) and same parity. Due to small number of levels in every unique nucleus, we have used all $2^+, 4^+$ levels for even mass nuclei and all $\frac{1}{2}^+, \frac{3}{2}^+, \frac{5}{2}^+$ levels for odd mass nuclei in the region $\leq 5\ Mev$ (which so called low-lying part). Firstly we include the number of the levels below $E$ and write it as [3]

$$N(E) = \int_0^E \rho(E)dE = e^{(\frac{E-E_0}{T})} - e^{-\frac{E_0}{T}} + N_0 \qquad , \qquad (4.1)$$

$N_0$ Introduces the number of levels with energies less than zero and it must be assume as zero [3]. The best fit to $N(E) (\equiv F(E))$ would be obtained if a correct set of energies is produced by means of [3]

$$E'_i = E_{min} + \frac{F(E_i) - F(E_{min})}{F(E_{max}) - F(E_{min})} (E_{max} - E_{min}), \qquad (4.2)$$

Both $E_{max}$ and $E_{min}$ remain unchanged with this transformation. These transformed energies should now display on average a constant level density. The spacing used in the determination of NNSD distributions are given by [3]

$$S_i = E'_{i+1} - E'_i \qquad , \qquad s_i = \frac{S_i}{D} \qquad (4.3)$$

where $D$ is the average of the spacing between corrected energy levels. Distribution $P(s)$ will be in such a way in which $P(s)ds$ is the probability for the $s_i$ to lie within the infinitesimal interval $[s, s+ds]$. These $s_i$ regarded as $X_i$ in Eq.(2.4) and then in Eq.(2.2) to estimate the density function with KD-based method. Also in every sequence, we have chosen the best value of $h$ (bin sizes) while minimize the ISE measure.

- **Spectral statistics of nuclei correspond to three dynamical symmetry limits of IBM and nuclei correspond to oblate and prolate classification in BMM**

This model has three dynamical symmetry limits while the classifications of different nucleus in these three dynamical symmetry limits have handled with their special dynamical properties and also their $R_{4/2} = \mathrm{E}^1_{4^+} / \mathrm{E}^1_{2^+}$ ratios. The nucleus correspond to these classification are listed in Table1 (the sequences prepared by $2^+$ and $4^+$ levels of these nuclei).

| Sequences | Nuclei |
|---|---|
| U(5) | $^{98}$Mo, $^{100}$Mo, $^{108}$Cd, $^{112}$Cd, $^{114}$Cd, $^{110}$Cd, $^{116}$Cd, $^{118}$Cd, $^{118}$Te, $^{120}$Te, $^{122}$Te, $^{124}$Te, $^{126}$Te, $^{112}$Sn, $^{114}$Sn, $^{134}$Xe, $^{154}$Dy,… |
| O(6) | $^{56}$Fe, $^{78}$Ge, $^{80}$Se, $^{130}$Ba, $^{132}$Ba, $^{132}$Ce, $^{134}$Ce, $^{196}$Hg, $^{194}$Pt, $^{196}$Pt, $^{198}$Pt, $^{198}$Hg,… |
| SU(3) | $^{166}$Er, $^{176}$Hf, $^{180}$W, $^{168}$Yb, $^{174}$Hf, $^{160}$Dy, $^{230}$Th, $^{184}$W, $^{232}$Th, $^{182}$W, $^{232}$U, $^{178}$Hf, $^{170}$Yb, $^{162}$Dy, $^{234}$U, $^{164}$Dy, $^{172}$Yb, $^{240}$Pu, $^{168}$Er, $^{170}$Er, $^{246}$Cm,… |

Table [1]. Nuclei correspond to three dynamical symmetry limits of IBM as have introduced in [10-11].

To compare the spectral statistics of these sequences with each other, we have evaluated the KLD measure related to GOE in all of them while the smaller distances reveal the more chaoticity degrees of systems. Table2 represents the ML estimated values for Brody distribution and also KLD measures display the distances of KD-based function to Wigner (chaotic) limit.

| statistical criterions \ Sequence | Nuclei with U(5) symmetry limit | Nuclei with SU(3) symmetry limit | Nuclei with O(6) symmetry limit |
|---|---|---|---|
| $\langle q \rangle$ The parameter of Brody distribution | $0.33 \pm 0.145$ | $0.57 \pm 0.211$ | $0.48 \pm 0.293$ |
| $\langle KLD \rangle$ | $1.922 \pm 0.112$ | $1.688 \pm 0.108$ | $1.741 \pm 0.144$ |

Table[2]. ML estimated values for Brody distribution and also KLD measure while display distances to GOE in different sequences. The uncertainties evaluated by IAE method.

The KLD measures confirm the previous statistical behavior while consider the more regular dynamics for U(5) dynamical symmetry limits in compare to other three dynamical symmetries of IBM and also, the rotational limit (SU(3) limit) represent the most chaotic dynamics. Also the noticeable reductions in the uncertainties of estimated density function (the uncertainties have evaluated with IAE) have been occurred, then, we can conclude, the KD-based function yield the closer density function to real and exact distribution of every sequences. Also, the KLD measures can regard as an appropriate criterion to explore the accuracy of different estimation methods in describing exact spectral statistics. On the other hand, as have been obtained in [11], the ML-based estimated value and density function exhibit more regularity in compare to other parametric estimation method in different sequence. To investigate the exact statistics of different sequences and compare the ML-estimated distribution with real distribution, the KLD values (while measure the distances to Poisson limit) have evaluated where listed in Table3. The KLD measures (for KD-based density function) confirm the more regularity even more than predicted by ML estimated values and therefore consider regular dynamics for nuclear systems more than predicted by other statistical approaches. Also the ML estimated values yield the closer distances to KD-based result and then estimate the closer one to exact distribution of every sequence.

| KLD \ Sequence | Nuclei with U(5) symmetry limit | Nuclei with SU(3) symmetry limit | Nuclei with O(6) symmetry limit |
|---|---|---|---|
| KD-based function | 1.312 | 1.515 | 2.041 |
| ML estimated function | 1.677 | 1.991 | 2.945 |
| LSF estimated function | 2.312 | 3.105 | 3.919 |

TABLE[3]. KLD measures while display distances to Poisson limit in different sequences for three estimation method in different sequences.

Also to compare the precision of KD-based density function in other sequences, we have studied the spectral statistics of nuclei correspond to oblate and prolate classification in the BMM framework while have been handled with LSF-based method and Abul-Magd distribution [33]. To investigate the statistical properties of these systems with the non-parametric estimation (KD-based), sequences were prepared by $2^+$ levels of nuclei listed in Table4 (similar to [33]).

| | |
|---|---|
| **Oblate** | $^{28}$Si, $^{26}$Mg, $^{72}$Se, $^{116}$Cd, $^{74}$Se, $^{76}$Se, $^{68}$Ge, $^{70}$Ge, $^{72}$Ge, $^{74}$Ge, $^{66}$Zn, $^{188}$Pt, $^{190}$Pt, $^{192}$Pt, $^{194}$Pt, $^{196}$Pt, $^{198}$Pt, $^{200}$Pt, $^{140}$Sm, $^{192}$Hg, $^{196}$Hg, $^{198}$Hg, $^{200}$Hg, $^{124}$Te, $^{62}$Ni, $^{202}$Hg, $^{204}$Hg, $^{204}$Pb, $^{206}$Pb, $^{214}$Po |
| **Prolate** | $^{150}$Gd, $^{152}$Gd, $^{154}$Gd, $^{188}$Os, $^{192}$Os, $^{228}$Ra, $^{58}$Fe, $^{108}$Pd, $^{146}$Ba, $^{148}$Nd, $^{150}$Sm, $^{154}$Dy, $^{160}$Yb, $^{228}$Th, $^{230}$Th, $^{232}$Th, $^{232}$U, $^{60}$Fe, $^{62}$Zn, $^{64}$Zn, $^{96}$Zr, $^{100}$Zr, $^{110}$Pd, $^{122}$Xe, $^{156}$Dy, $^{234}$U, $^{240}$Pu, $^{246}$Cm, $^{184}$W, $^{100}$Mo, $^{152}$Sm, $^{164}$Yb, $^{182}$W, $^{180}$Pt, $^{182}$Pt, $^{250}$Cf, $^{160}$Dy, $^{162}$Er, $^{166}$Yb, $^{106}$Ru, $^{162}$Dy, $^{164}$Dy, $^{168}$Er, $^{172}$Hf, $^{174}$Hf, $^{178}$Hf, $^{170}$Yb, $^{170}$Er, $^{172}$Yb, $^{28}$Mg, $^{24}$Mg, $^{76}$Kr, $^{22}$Ne |

Table [4]. Nuclei correspond to prolate and oblate classification of BMM framework as have introduced in [33].

As have been exhibited in [33], the oblate nuclei explore the deviation to regular dynamics in compare to prolate one while the KLD measures confirm the similar statistical properties with reduction in the uncertainties (Table5).

| Sequence<br>statistical criterions | Oblate nuclei | Prolate nuclei |
|---|---|---|
| $\langle q \rangle$ parameter of Abul-Magd distribution | $0.56 \pm 0.247$ | $0.64 \pm 0.358$ |
| $\langle KLD \rangle$ | $1.695 \pm 0.128$ | $1.311 \pm 0.144$ |

Table[5]. ML estimated values for Abul-Magd distribution and also KLD measure while display distances to GOE in different sequences. The uncertainties evaluated by IAE method.

Figures2,3 represent the NNSDs for this five sequences (three dynamical symmetry limits of IBM and also oblate and prolate nuclei) while displayed by histogram and KDE methods respectively.

### 4.2. Investigation of pairing effect on spectral statistics

As have mentioned in Section3, the pairing force regarded as the main reason for regular dynamics in the low-lying region of nuclear spectra. In order to confirm this feature of nuclear force, we have prepared sequences by different nuclei (will introduce in the following) to investigate the effect of pairing in different mass regions, the effect of pair structure (particle-particle or hole-hole) and also pair type (proton-proton or neutron-neutron) on spectral statistics.

#### 4-2-1) Paired systems(even mass nucleus) versus unpaired systems (odd-mass nucleus)

To reveal the pairing effects in general, we have selected two sequences by nucleus tabulated in Table3 as paired and unpaired ones.

| Sequences | Nuclei |
|---|---|
| **Even-Even mass nuclei** | $^{44}$Ca, $^{46}$Ca, $^{48}$Ti, $^{50}$Ti, $^{50}$Cr, $^{52}$Cr, $^{54}$Cr, $^{58}$Fe, $^{66}$Zn, $^{68}$Zn, $^{68}$Ge, $^{70}$Ge, $^{72}$Ge, $^{74}$Ge, $^{76}$Se, $^{78}$Se, $^{82}$Kr, $^{84}$Kr, $^{90}$Zr, $^{92}$Zr, $^{96}$Mo, $^{98}$Mo, $^{102}$Ru, $^{104}$Ru, $^{106}$Ru, $^{102}$Pd, $^{106}$Pd, $^{108}$Pd, $^{110}$Pd, $^{110}$Cd, $^{112}$Cd, $^{116}$Cd, $^{118}$Sn, $^{120}$Sn, $^{122}$Xe, $^{124}$Te, $^{126}$Te, $^{150}$Gd, $^{152}$Gd, $^{154}$Gd, $^{164}$Yb, $^{182}$W, $^{180}$Pt, $^{182}$Pt, $^{188}$Os, $^{190}$Os, $^{192}$Os, $^{192}$Hg, $^{196}$Hg, $^{198}$Hg, $^{200}$Hg |
| **Odd-mass nuclei** | $^{43}$Ca, $^{47}$Ti, $^{49}$Ti, $^{53}$Cr, $^{57}$Fe, $^{61}$Ni, $^{67}$Zn, $^{73}$Ge, $^{81}$Kr, $^{77}$Se, $^{87}$Sr, $^{91}$Zr, $^{95}$Mo, $^{99}$Ru, $^{105}$Pd, $^{111}$Cd, $^{117}$Sn, $^{125}$Te, $^{131}$Xe, $^{137}$Ba, $^{141}$Ce, $^{143}$Nd, $^{145}$Pm, $^{149}$Sm, $^{151}$Eu, $^{155}$Gd, $^{159}$Tb, $^{161}$Dy, $^{163}$Ho, $^{167}$Er, $^{169}$Tm, $^{173}$Lu, $^{179}$Hf, $^{183}$W, $^{187}$Re, $^{189}$Os, $^{191}$Ir, $^{195}$Pt, $^{199}$Hg, $^{197}$Au |

Table[6]. Even and Odd mass nuclei which were used to prepare two paired and unpaired sequences.

As have been presents theoretically in [57-59], the intensity of pairing interaction in the Hamiltonian of Even-Even mass nuclei have considered with stronger intensity in compare to the Hamiltonian of odd-mass ones. On the other hand, pairing force have introduced as the main reason of regular dynamics in spectral statistics of nuclear systems, therefore, we can predict the deviation to regularity in Even-even nuclei while as it is displayed in Table7 and Figure 4, even-even mass nuclei (where pairing force is dominant) has bigger distance to GOE, therefore they have more regular dynamics in compare to odd-mass nuclei (unpaired systems). Also, the ML-estimated values for Brody distribution reveal this statistical behavior (small q, represent the more regular dynamics).

| Sequence / statistical criterion | Even-mass nuclei | Odd-mass nuclei |
|---|---|---|
| $\langle q \rangle$ The parameter of Brody distribution | 0.19±0.3758 | 0.43±0.3122 |
| $\langle KLD \rangle$ Related to GOE | 1.5208±0.1247 | 0.9580±0.1351 |

Table[7]. ML estimated values for Brody distribution and also KLD measure represent distances to GOE limit for Even-Even and Odd-mass nuclei which demonstrates regular dynamic for Even-Even systems in compared to Odd-mass systems. The uncertainties evaluated by IAE method.

The reductions in the uncertainties (while evaluated by ISE) confirm the more accuracy with KD-based estimated density function in compare to parametric estimation methods.

### 4-2-2) Pairing in different mass regions

As have mentioned in introduction, we have investigated the pairing effect in different mass region to exhibit the effect of closed shell on statistical properties of nuclear systems. To handle this comparison, we have utilized the shell model configuration and classified the Even-Even mass nucleus (some of nuclei where have used in previous part) in the two region, nuclei with $50 \leq A \leq 100$ (N(or Z): 20 to 50) and the second region with nuclei located in the $150 \leq A \leq 210$ mass region (N(or Z): 50 to 82) which are listed in Table8.

| Sequences | Nuclei |
|---|---|
| **Even-mass nuclei with N(or Z): 20 to 50** | $^{50}$Cr, $^{52}$Cr, $^{54}$Cr, $^{58}$Fe, $^{66}$Zn, $^{68}$Zn, $^{68}$Ge, $^{70}$Ge, $^{72}$Ge, $^{74}$Ge, $^{76}$Se, $^{78}$Se, $^{82}$Kr, $^{84}$Kr, $^{90}$Zr, $^{92}$Zr, $^{96}$Mo, $^{98}$Mo |
| **Even-mass nuclei with N(or Z): 50 to 82** | $^{150}$Gd, $^{152}$Gd, $^{154}$Gd, $^{164}$Yb, $^{182}$W, $^{180}$Pt, $^{182}$Pt, $^{188}$Os, $^{190}$Os, $^{192}$Os, $^{192}$Hg, $^{196}$Hg, $^{198}$Hg, $^{200}$Hg |

Table[8]. Even mass nuclei in two different mass regions.

The KLD measures while display the distances of KD-based estimated density function to Wigner or chaotic limit, confirm more regularity for heavier nuclei in compared to lighter ones which reveal theoretical predictions about chaotic dynamics of lighter nuclei [3,11] (similar to ML estimated values for the parameter of Brody distribution where estimate very closer to Poisson limit ($q \to 0$) for heavier nucleus). Figure 5 and Table 9 represent this comparison between these two sequences.

| Sequence / statistical criterion | **Even-mass nuclei with N(or Z): 20 to 50** | **Even-mass nuclei with N(or Z): 50 to 82** |
|---|---|---|
| $\langle q \rangle$ The parameter of Brody distribution | 0.63±0.2984 | 0.39±0.1955 |
| $\langle KLD \rangle$ Related to GOE | 1.4003±0.1003 | 1.6691±0.0953 |

Table[9]. ML estimated values for Brody distribution and also KLD measure represent distances to GOE limit for sequences introduced in Table8 which demonstrates regular dynamic for heavier even-mass nucleus in compared to lighter ones. The uncertainties evaluated by IAE method.

Similar to previous results, the reduction of uncertainty with KLD-based estimated function and closer approach to the exact distribution of these sequences are verified with IAE-based values.

### 4-2-3) The effect of identical pairs on nuclear spectra

The spectral statistics due to the effect of pair type in nuclear systems have been studied theoretically in [58-59] where suggest a regular dynamics for systems with neutron-neutron pairs. To carry out a similar analysis with pure experimental data, we have prepared two sequences of nuclei with different pair types: nuclei with proton-proton pairs (with full neutron energy levels in the configuration of shell model) and the second one by nuclei with neutron-neutron pairs (with full proton energy levels in the configuration of shell model) as is given in Table10.

| Sequences | Nuclei |
|---|---|
| Nuclei with proton-proton pairs | $^{50}$Ti, $^{70}$Zn, $^{74}$Se, $^{86}$Kr, $^{92}$Mo, $^{98}$Mo, $^{118}$Te, $^{140}$Ce, $^{168}$Er |
| Nuclei with neutron-neutron pairs | $^{34}$S, $^{42}$Ca, $^{64}$Ni, $^{76}$Se, $^{74}$Ge, $^{140}$Ba, $^{156}$Gd, $^{43}$Ca, $^{158}$Dy, $^{172}$Yb |

Table[10]. Nucleus with different pair types, proton-proton and neutron-neutron pairs.

The ML estimated values for Brody distribution and also the KD-based estimated density function and the corresponding KLD values while represent distances to chaotic (Wigner) limit, exhibit more regularity for nuclei with neutron-neutron pairs in compared to others. On the other hand, nuclei with proton-proton pairs which Coulomb force reduces the effect of pairing force in nuclear energy spectra, show more chaoticity. Therefore, we can conclude, pairing and Coulomb forces are competing with each others in order to dominate the regularity or chaoticity characteristics of nuclear spectra, respectively. Table11 and Figure 7 display these results where remarkable reductions in uncertainties are obvious.

| Sequence / statistical criterion | Nuclei with proton-proton pairs | Nuclei with neutron-neutron pairs |
|---|---|---|
| $\langle q \rangle$ The parameter of Brody distribution | $0.54 \pm 0.2977$ | $0.32 \pm 0.2173$ |
| $\langle KLD \rangle$ | $1.2576 \pm 0.0811$ | $1.6484 \pm 0.114$ |

Table[11]. ML estimated values for Brody distribution and also KLD measure represent distances to GOE limit for sequences introduced in Table10 which demonstrates regular dynamic for nuclei with proton-proton pairs in compared with nuclei with neutron-neutron pairs. The uncertainties evaluated by IAE method.

### 4-2-4) The effect of identical holes on nuclear spectra

Similar to previous subsection and in order to explore another statistical property due to pairing effect in some nuclei with unclosed shells, we have used the shell model configuration of levels and classified nucleus in two groups, nuclei whose protons occupy levels (in shell model scheme) completely but the neutron levels have some empty states and the second group nuclei with vice versa condition. We have prepared our sequences by nuclei listed in Table12.

| Sequences | Nuclei |
|---|---|
| **Nuclei with holes in neutron levels** | $^{46}$Ca, $^{70}$Ge, $^{82}$Se, $^{94}$Zr, $^{112}$Sn, $^{168}$Yb, $^{206}$Pb |
| **Nuclei with holes in proton levels** | $^{38}$Ar, $^{58}$Fe, $^{88}$Sr, $^{114}$Cd, $^{144}$Sm, $^{154}$Gd, $^{168}$Er |

Table[12]. Nucleus with unfilled neutron levels (holes in neutron levels) and also nucleus with unfilled proton levels.

The resultant values (ML estimated values for the parameter of Brody distribution and further the KLD values while explain the distances of KD-based estimated density function) exhibit more regular dynamics for nuclei with unfilled neutron levels. It's necessary to say, we can't regard this deviation to regularity due to the only pairing effect where the pairing interaction is dominant in the both sequence but as have mentioned in previous part, we can exhibit this regularity as the effect of competition between pairing and Coulomb forces while the stronger Coulomb force in nuclei with unfilled proton levels yield a more chaotic dynamics. Also, the increasing the accuracy with KD-based estimation in compares to ML estimated values and corresponding distribution is reveal similar to other sequences.

| Sequence \ statistical criterion | Nuclei with holes in neutron levels | Nuclei with holes in proton levels |
|---|---|---|
| $\langle q \rangle$ The parameter of Brody distribution | $0.40 \pm 0.1962$ | $0.54 \pm 0.2385$ |
| $\langle KLD \rangle$ Related to GOE | $2.0611 \pm 0.0794$ | $1.0375 \pm 0.1081$ |

Table[13]. ML estimated values for Brody distribution and also KLD measure represent distances to GOE limit for sequences introduced in Table12 which demonstrates regular dynamic for nuclei with unfilled neutron levels in compared to nuclei with holes in proton levels. The uncertainties evaluated by IAE method.

## 5. Conclusion and Summary

In the present paper, KDE method as a non-parametric estimation method is utilized in investigating the statistical properties of nuclear spectra in NNSD framework. The regular or chaotic dynamics of every spectra exhibited by KLD measures while represent the closer distances to Poisson or Wigner limits, respectively. Using KDE method, we have estimated the density function in sequences of nuclei correspond to three dynamical symmetry limits of IBM, nuclei correspond to oblate and prolate classification in BMM framework. Also we have investigated the pairing effect on spectral statistics of sequences prepared by nuclei with different types of pairs. In all cases, the KDE estimated density functions have minimum uncertainties in compare to those estimated by MLE method while evaluated by ISE method. Therefore, the KD-based estimated functions prepare the closer density function to real distribution of every sequence. Also, KD-based estimated functions explore more regularity for different nuclear systems even more than the prediction of ML-based estimation. Furthermore, the deviation to regular dynamics due to pairing interaction is confirmed in different sequences. This approach can be considered as a very exact method in spectral investigation while can exhibit statistical properties independent of special distributions which we will attempt in the next papers.

## Figure caption

Figure1. Casten triangle [31-32] which describe three symmetry limits of IBM, as have displayed for every limit, regular spectra for U(5) limit in compared to others is obvious.

Figure2. NNSDs (based on histogram) for three dynamical symmetry of IBM respectively (in the first line) and also nuclei correspond to oblate and prolate classification of BMM (in the second line). Solid line, dashed line and dotted line represent Histogram, Poisson and GOE curves respectively.

Figure3(color online). Similar to Figure2, based on KDE methods for 5 sequences.

Figure (color online). NNSDs for Even-Even mass and odd-mass nuclei based on KDE method. Solid line, dashed line and dotted line represent Histogram, Poisson and GOE curves respectively.

Figure5 (color online). NNSDs for Even-Even mass nucleus in two mass regions, nuclei with $(N(or Z))$: 20~50 and also nuclei with $(N(or Z))$: 50~82 based on KDE method. Solid line, dashed line and dotted line represent Histogram, Poisson and GOE curves respectively.

Figure6 (color online). NNSDs for nuclei with proton-proton pairs and nuclei with neutron-neutron pairs based on KDE method. Solid line, dashed line and dotted line represent Histogram, Poisson and GOE curves respectively.

Figure7 (color online). NNSDs for nuclei with unfilled neutron levels (holes in neutron levels) and nuclei with unfilled proton levels (holes in proton levels) based on KDE method. Solid line, dashed line and dotted line represent Histogram, Poisson and GOE curves respectively.

Figure1.

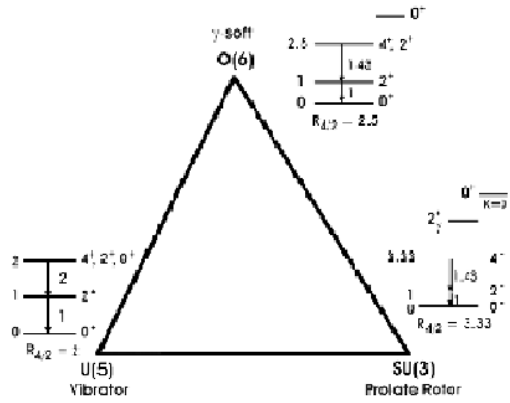

Figure2.

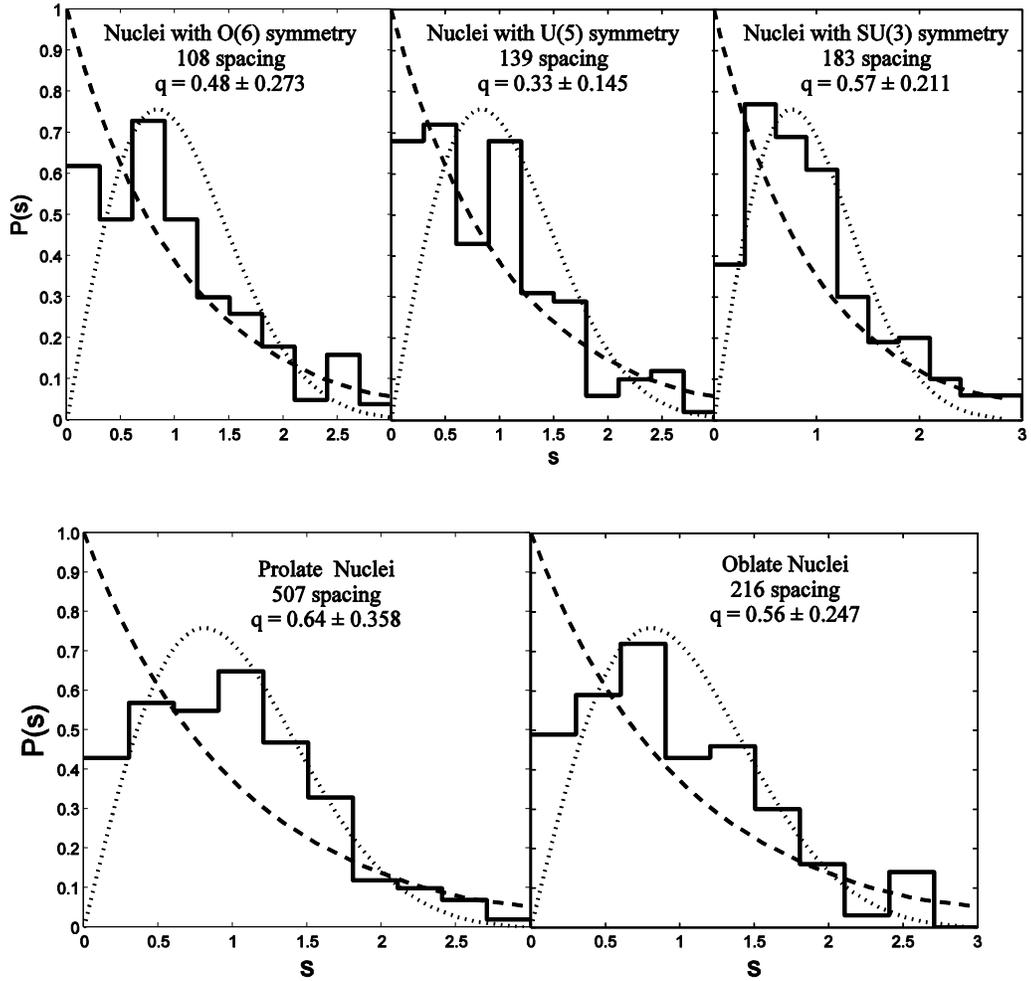

Figure3.

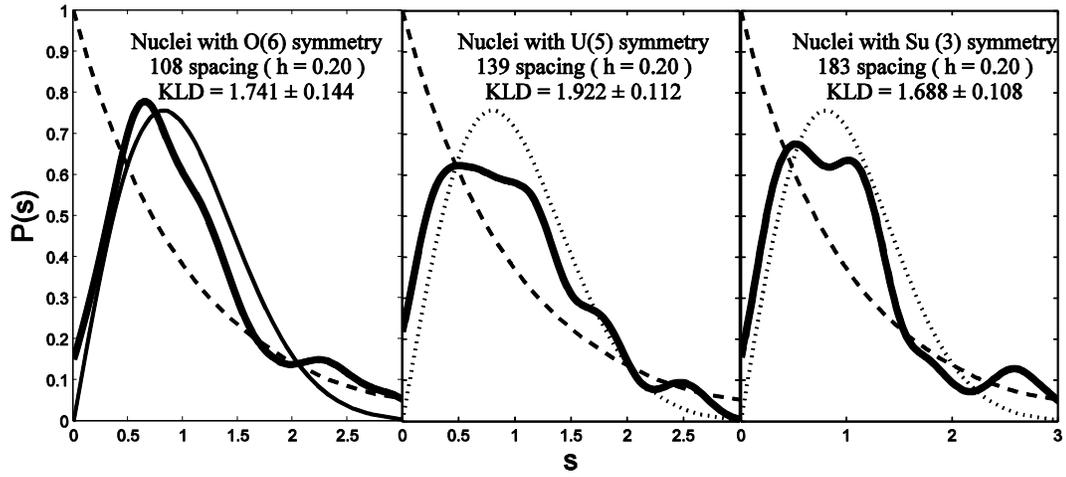

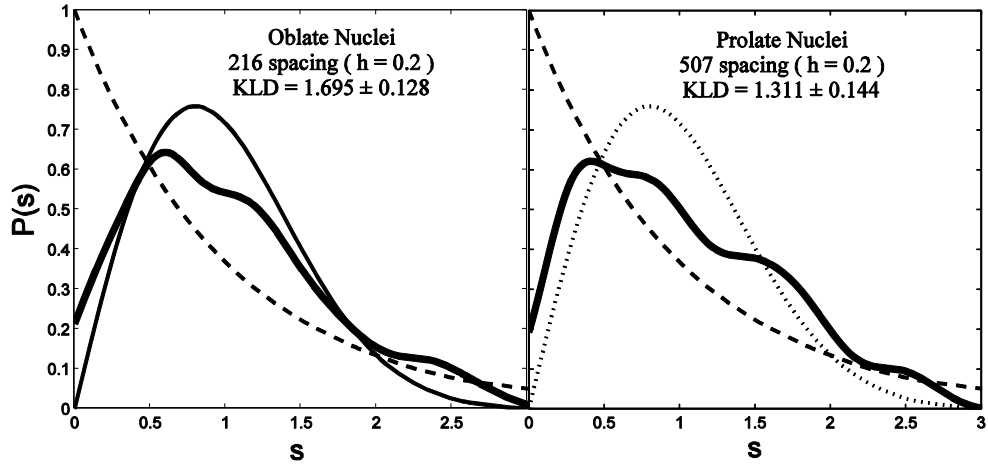

Figure4.

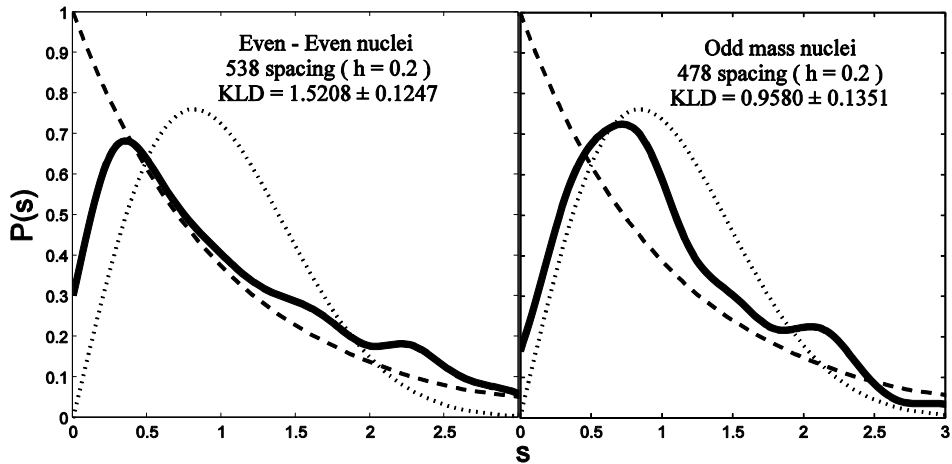

Figure5.

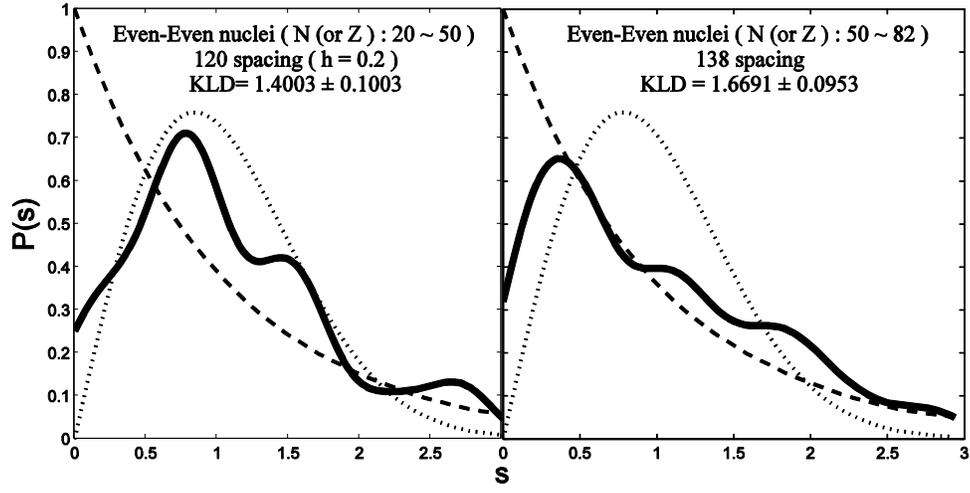

Figure6.

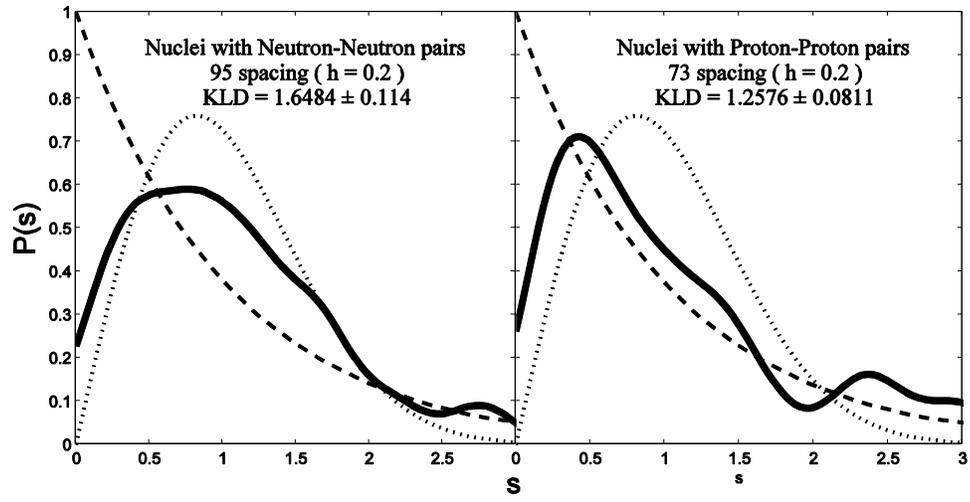

Figure7.

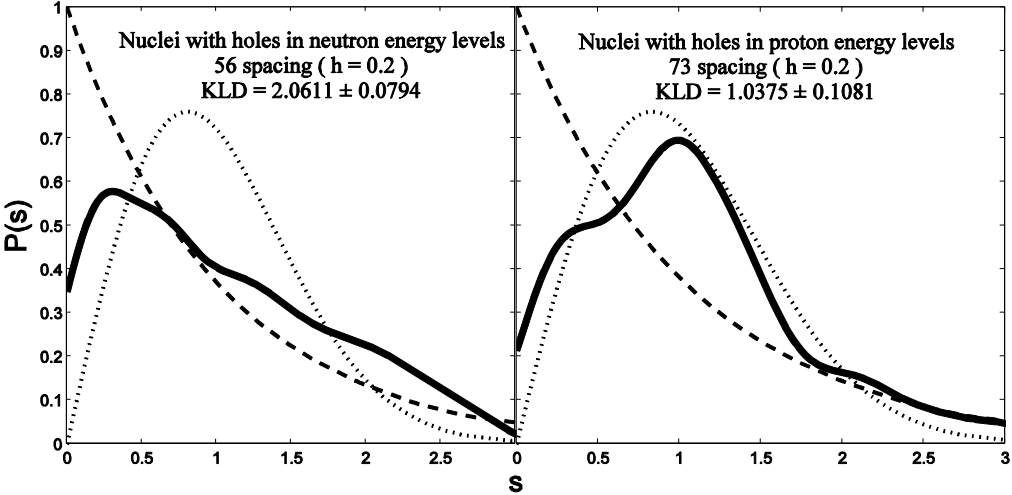